\documentclass{epl}

\newcommand{\gammastar}{\Gamma^* }

\title{Clustering in gravitating N-body systems}
\author{
	M. Bottaccio\inst{1,2} \and
	A. Amici\inst{1} \and
	P. Miocchi\inst{1} \and
	R. Capuzzo Dolcetta\inst{1} \and
	M. Montuori\inst{2} \and
	L. Pietronero\inst{1,2}
}
\institute{
  \inst{1}  Physics Department, University of Rome ``La Sapienza'', Italy, \\
  \inst{2} INFM, Research Unit of Roma ``La Sapienza'', Italy.
}
\pacs{05.20.-y}{Classical statistical mechanics}
\pacs{45.50.Jf}{Few- and many-body systems}
\pacs{45.50.-j}{Dynamics and kinematics of a particle and a system of particles}
\shortauthor{M. Bottaccio \etal}
\begin{document}

\maketitle

\begin{abstract}
We study  gravitational clustering of mass points in three dimensions with
random initial positions and periodic boundary conditions (no expansion)
by numerical simulations.
Correlation properties are well defined in the system and
 a sort of thermodynamic limit can be defined for the transient regime of clustering.
Structure formation proceeds along two paths: (i) fluid-like evolution of 
density perturbations at large scales and (ii) shift of the granular
(non fluid) properties from small to large scales.
The latter mechanism finally dominates at all scales and it is responsible 
for the self-similar characteristics of the clustering.
\end{abstract}

One of the fundamental challenges of modern cosmology
is the understanding of the formation of the structures in the
Universe.
These structures consist of clusters of galaxies and show  complex 
properties extended to very large scales ~\cite{sylos:montuori:pietro:98}.
 Usually 
the simulations and the models aimed at the understanding 
of these structures are based  
on three essential elements: (i) the dynamics under the effect
 of the gravitational forces; (ii) some particular type of initial conditions;
(iii) a model for the Hubble expansion ~\cite{padman:93,saslaw:00}.
In addition simulations are usually run up to a  time which is supposed to 
represent the present state.

Here we would like to take inspiration from these studies and  
formulate the problem of clustering by gravity in the perspective of the 
statistical physics of dynamical systems.
So we will single out the role of each individual effect at the expense of a 
loss in realism.
We try therefore to  identify simple fundamental mechanisms   which
can be studied in great detail and followed up to their asymptotic state.
As a first problem we consider 
the simple, basic question: how does a random distribution of
point masses evolve under gravity?
The comparison with the expanding case may then allow us to identify 
the specific role of this effect. Simulations similar to ours, 
but in a cosmological context,
 were performed long ago, e.g., by  Itoh et al.~\cite{itoh:saslaw:88}.
However, we are going to see that the general problematic
we consider and the final interpretation will be substantially different.

The main results are:
(i) the existence of a well defined thermodynamic limit for the
transient regime;
(ii) the nature of the clustering process arising from the shift 
of the granular (non fluid-like) characteristics from small to large scales.
(iii) the evolution of correlations shows self similar characteristics.

Self-gravitating systems
have  been studied in various perspectives since a long time. However,
due to the peculiarities of the $1/r$ 
potential, most of the concepts and methods of standard statistical 
physics are problematic~\cite{hertel:thirring:71,kandrup:80,torcini:antoni:99,devega:sanchez:01:I}.
Indeed, the usual approach of statistical mechanics cannot be applied
to a system of many point particles interacting by
the Newtonian potential, because of
(i) the long range nature of the $1/r$ potential
and of (ii) the divergence at the origin~\cite{padman:90}.
In particular, energy is not extensive, therefore the thermodynamic properties 
cannot be uniquely
defined as the microcanonical and canonical ensembles are not equivalent.


In order to simulate an infinite system, 
 $N$ particles at rest are placed at random into a cube of side $L$ with 
periodic boundary conditions.
Every particle in the simulation box interacts with 
all the other particles and with the 
periodic ``replicas'' of the whole system. 
The Hamiltonian for a system of $N$ identical particles
of mass $m$ is:
\begin{equation}
H
=
\sum_i {
	\frac{
		{\bf p}^2_i
	} {
		2m
	}
}
+
\sum_{\left<ij\right>} {
	m \, \Phi^{L}
	\left(
		{\bf r}_i - {\bf r}_j
	\right)
}
,
\label{eq:generic_H}
\end{equation}
where the second summation runs over all pairs.
The interaction potential which accounts for both the pair contribution
and that of the replicas, is:
\begin{equation}
\Phi^{L}
	\left(
		{\bf r}
	\right)
=
\sum_{\bf n} {
\left[
-
	\frac{
		Gm
	} {
		| \, {\bf r} + L{\bf n} \, |
	}
+
	\int_V
	\frac{
		Gm/L^3
	} {
		| \, {\bf r'} + L{\bf n} \, |
	} \rm{d}{\bf r'}
\right]
},
\label{pair_potential}
\end{equation}
where $G$ is the gravitational constant,
$V$ is the box domain and ${\bf n}$ runs over all three dimensional vectors
with integer components.
In the above integral the first term is the usual Newtonian potential
due to a particle with mass $m$ placed in the replica 
specified by {\bf n}, the second
term is introduced  to remove the long range divergence of the potential.
It can be seen as the potential generated by a repulsive mass $-m$
smeared uniformly over the volume of the same replica. The summation over 
the replicas is given by the summation of {\bf n}. The system is 
individuated by ${\bf n}=0$. 
Such definition for $\Phi^{L}$ gives the expected expression for the force 
due to an infinite periodic system.
Periodic boundary conditions have two major advantages with respect to
free boundaries, when simulating an infinite system 
with a finite N-body representation, namely: (i) they 
avoid having a preferential point 
in the simulation and (ii) the total number of particles inside the 
cube of size $L$ is constant, so particles cannot ``evaporate'', as
 would happen in an open system~\cite{chandra:60}.
Furthermore they ensure that both terms in Eq. (\ref{eq:generic_H})
grow linearly with $N$ for constant density. 
In order to simulate the evolution
 of a large number of 
particles the use of an efficient method for the  evaluation of forces 
is necessary.
For our simulations we used a tree-like algorithm of the kind firstly
proposed by Barnes et al.~\cite{hut:barnes:86}.
In this algorithm, the force acting on a particle due to its neighbors 
is evaluated exactly, while that due to distant particles
is evaluated by means of a multipolar expansion.

The nature of the gravitational potential requires a
time integrator, which is both flexible and
efficient. Actually, the short range divergence can produce 
fast changes in the dynamical variables of a particle,
whose  motion should be therefore followed with 
 a very short time step.
Nevertheless, most of the 
particles may have a regular and smooth motion 
which doesn't need  such a careful integration. 
The solution adopted is thus an algorithm, based on a variable and
individual time-step.
In order to avoid time steps approaching to zero when 
a very close encounter between two 
particles occurs, a smoothing in the potential 
is introduced at very small scales. 
This, often-used, technique introduces of course 
an artificial length scale in the 
simulations. In order to save the Newtonian behavior of the potential,
such a scale 
must be taken much smaller than the average distance 
between nearest neighbors 
$\lambda$.
In the simulations presented here, the smoothing scale is approximately
$1/32 \lambda$. We have checked on some simulations
with $2^{12}$ particles that using a smaller
smoothing length does not influence our results; however it slows down
even more the code. 

The simulations presented in this paper were performed using  the tree-code 
described in ~\cite{miocchi:98,dolcetta:miocchi:98}  which 
we adapted to represent an infinite system by means of periodic boundary 
conditions according to  Ewald prescription (see e.g. \cite{hernquist:bouchet:suto:91}).

A small smoothing scale and the absence of cosmological expansion,
together with the requirement of high accuracy imply computations
heavier than the usual cosmological ones. This is because
we consider the generation of structures from a strongly non linear dynamics
starting from random initial positions.
This will limit the number of particles $N$ in our systems,
 but all our results will be analyzed with respect to the dependence on $N$.
Our simulations are presented using arbitrary units
for mass ($m_u$), length ($r_u$) and time ($t_u$).
The Newtonian equations depend on a  single constant
with physical dimensions ($G$), therefore 
we can set arbitrarily two of the units and
the third is uniquely defined according to $r_u^3 t_u^{-2} m_u^{-1}=G$.

\begin{figure}
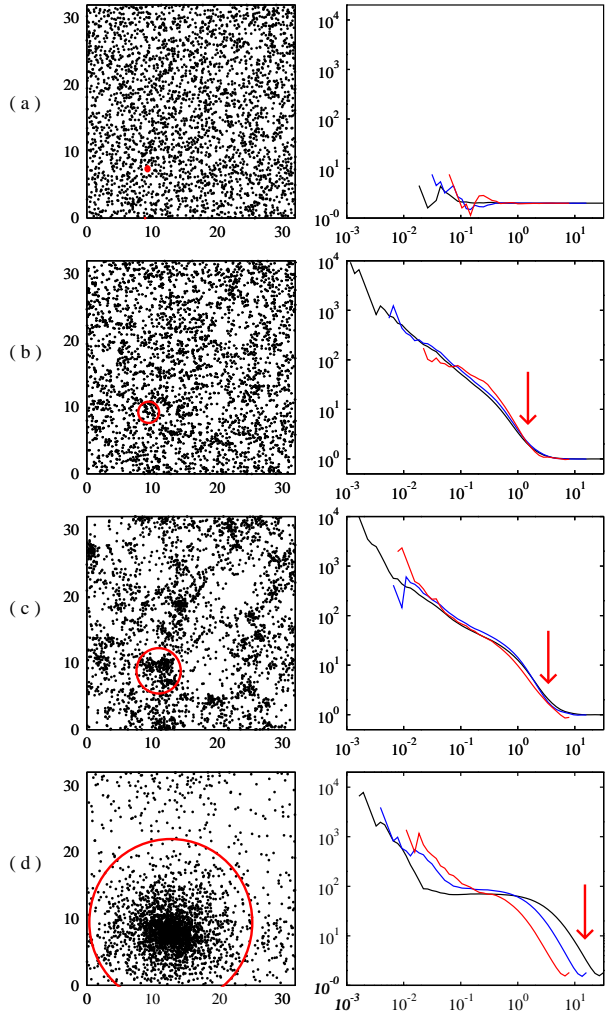

\onefigure[width=8.5cm]{fig1.eps}
\caption{Left: projection onto the $xy$-plane at different times of a system
with $N=2^{15}$ particles in a simulation box with side $L= 32$. 
(a) $t= 0$; (b) $t= 0.6$; (c) $t= 1$; (d) $t= 3.4$. Only $1/4$ of the points
are shown for the sake of clarity. 
Right: Conditional average density of systems with $N=2^{9}$ (red), 
$N=2^{12}$ (blue), 
$N=2^{15}$ (black). Arrows mark the ``homogeneity scale''. Note 
that the structures
of the first three figures are size independent, while  the size of the system
 influences the final cluster.
}
\label{fig:projection}
\end{figure}

The left column of fig.~\ref{fig:projection}
shows the projection onto the $xy$-plane of
typical snapshots of a system with $N=2^{15}$ particles
at different times. All our simulations are performed with a constant
number density $n_0= N/L^3=1 \, r_u^{-3}$. With this choice, a typical
crossing time  $\tau= (m n_0 G)^{-1/2}$ corresponds to $1 t_u$.
On the right column we plot the conditional average
density\cite{coleman:pietronero:92} $\Gamma^*(r,\,t)$
of the system at the same time of the corresponding snapshot.
 $\Gamma^*(r,\,t)$ is the average
density in spheres of radius $r$ centered on a particle $i$ of
the system. It can be defined as:
\begin{equation}
\Gamma^*(r, \, t)
=
\frac{m}{\Omega(r) N}
\sum_{i\neq j}
\sum_{\bf n}
	\theta(r - |{\bf r}_i(t) - {\bf r}_j(t) + L{\bf n}|) 
\label{eq:gamma}
\end{equation}
where $\Omega(r) = 4\pi r^3 / 3$ and ${\bf r}_i(t)$,  ${\bf r}_j(t)$
are the positions at time $t$ of particles
 $i$ and $j$ respectively. 

In the eq.~\ref{eq:gamma}, particles
 $i$ and $j$ belong to the simulation box, and the  summation on the 
integer vectors ${\bf n}$ 
allow to take into account the replicas of the particle $j$. 
The three lines in fig.~\ref{fig:projection} correspond to
simulations with  $N=2^{9}$ (red), 
$N=2^{12}$ (blue), 
$N=2^{15}$ (black).
Initial conditions (a) are given by $N$ point masses located randomly 
and at rest into the simulation box. In the initial phase (b) the system 
develops clustered structures on small scales. Correspondingly
 the $\Gamma^*(r,\,t)$ acquires amplitude below the homogeneity scale 
$r_0$, 
defined by the condition $\Gamma^*(r_0,\,t)= 2 n_0$. 
The scale $r_0$ can be seen as a typical clustering scale; the reason for
such definition will be clearer with the definition of
$\overline{\xi}(r, \, t)$.
The red circle in the corresponding snapshot has radius $r_0$, and roughly
identifies the scale at which the fluctuations in the number density are of the same order of the average value $n_0$.

As long as the cluster sizes are much smaller than the size
of the box they continue merging and forming bigger and bigger structures (c).
Accordingly the homogeneity scale $r_0$ increases. Eventually, the typical
 size of the few remaining aggregates becomes comparable with
the box size 
 and they 
collapse into a single big cluster which contains almost all the matter (d).
In order to give a rough estimate the time scale for the formation of the
 final cluster, we can define a transition time $t_f$, such that 
$r_0(t_f)= L/4$.
For simulations with the same mass density, larger $N$
corresponds to larger box size $L$; 
the final cluster is therefore larger and $t_f$
is greater.
After the time $t_f$
the system reaches a genuine statistically-stationary state.
The accuracy of our simulations can be checked 
by energy conservation. It is a bit tricky to 
evaluate energy conservation,  because in an infinite system there is no 
absolute value for the potential. Then the 
 relative error in energy $\Delta E/E$ is not a meaningful
quantity. For this reason we
use the quantity $\Delta E(t)/\Delta K(t)$, the ratio
between the error in the total energy and the increase in kinetic
energy.
 Typically, when the final cluster forms
the energy conservation is within $1\%$ . 

In the first three snapshots of fig.~\ref{fig:projection}
the measured $\Gamma^*(r,\,t)$ function does not depend on the
number $N$ of particles in the simulation box. This is because
the characteristic size of the structures is much smaller than
the box size.
On the other hand, we can see 
in the last panel (d) 
that the $\Gamma^*(r,\,t)$ corresponding to
systems with different number of particles do not overlap
any more. Therefore, the size of the final cluster depends 
on the box size. 

\begin{figure}
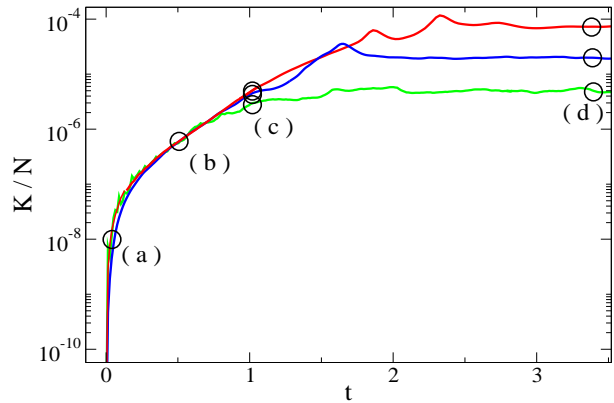

\onefigure[width=8cm]{fig2.eps}
\caption{
	Kinetic energy per particle for 
simulations with  $N=2^{9}$ (red), 
$N=2^{12}$ (blue), 
$N=2^{15}$ (black). Circles identify
the times used in fig.~\ref{fig:projection}. Note that 
the dynamics of the simulations becomes different only at 
large times, when the cluster size is comparable to the
box size.
}
\label{fig:kinetic}
\end{figure}
In figure~\ref{fig:kinetic} we show the kinetic
energy per particle $K(t)$  versus time 
for the three simulations.
We have marked the values of the kinetic energy
corresponding to the snapshots in fig.~\ref{fig:projection}.
Again, we can distinguish two regimes in the time evolution
of $K(t)$: (i) during the initial collapse phase
$K(t)$ show the same growth for the three simulations
while (ii) at a later time, the equilibrium value
depends strongly on the number $N$ of particles.
The breakdown of the Gibbs statistical treatment
is reflected here
by the fact that equilibrium properties
of the system, like $K(t\rightarrow \infty)$ and
$\Gamma^*(r, \, t \rightarrow \infty)$,
do not posses a well-defined thermodynamic limit.
In other words, they do not posses a finite
limit for $N \rightarrow \infty$ and
$L \rightarrow \infty$ with $N/L^3$ kept constant.
On the other hand,
the same quantities appear to be independent on the
size of the system for times smaller than  $t_f$.
This suggests that the transient phase, during which collapse
occurs, does posses properties with a well behaved limit
in the infinite system.
This implies that the transient clustering is a well defined
statistical properties of the system.

\begin{figure}
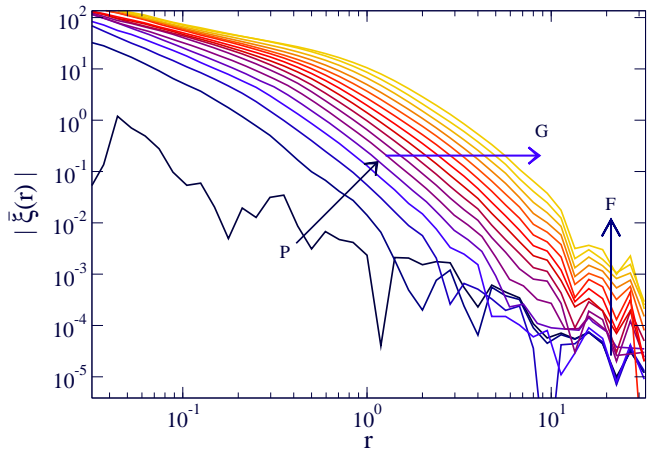

\onefigure[width=8.5cm]{fig3.eps}
\caption{
	Evolution of $| \overline{\xi}(r) |$ between $t= 0$ and $t= 1$, 
spaced by time intervals $\Delta t= 1/15$  for a system 
with $N = 2^{15}$ particles. Arrow $P$ identifies the initial phase
dominated by two particles interactions; $F$ is associated with 
fluid-like evolution at large scales,but the main dynamics for cluster
formation is due to the process indicated by $G$, which effectively
shifts the granularity from small scales to large scales.
}
\label{fig:xi}
\end{figure}

In order to characterize the very small density fluctuations
at the largest scales of the simulations, we make use the integrated
two-point correlation function
of density fluctuations
$\overline{\xi}(r, \, t) = \gammastar(r, \, t)/n_0 - 1$.
It measures the fluctuations above the average value $n_0$
 inside a sphere of radius $r$. 
Our definition of $r_0$ then corresponds to $|\overline{\xi}(r_0, \, t)|=1$,
which is a standard definition for the homogeneity scale~\cite{gaite:99} 
In fig. \ref{fig:xi} we show the absolute value
  $|\overline{\xi}(r, \, t)|$ 
versus $r$ for the system with $N=2^{15}$.
The lowest (black) line represents the correlation function
of the initial configuration, which scales as
$r^{-3/2}$ as expected for points distributed randomly.
The other lines  corresponds to times $t$ spaced
by 1/15 $t_u$.
At the largest scales of our simulations,
say for $r > 10 r_u$,
the time evolution of the correlation function appears to
be well described by the linear approximation
for the equation of a self-gravitating fluid \cite{jeans:28},
which predicts:
\begin{equation}
\overline{\xi}(r, \, t) = \overline{\xi}(r, \, 0) * 
	\cosh^2{\left( t/{\tau_l} \right)}
\label{eq:linear}
\end{equation}
with $\tau_l \approx (4\pi G\rho_0)^{-\frac{1}{2}} = 1/\sqrt{4 \pi} \, t_u$.
Eq.~\ref{eq:linear} describes the linear regime in which the
time evolution of the system consists solely in an
exponential amplification of the initial density
fluctuations. Such behavior is indicated in the figure, by
the upward arrow labeled by $F$.
It is worth noting that in this regime, overdensities
increase while underdensities actually decrease;
however, in fig.~\ref{fig:xi}
we have plotted the absolute value of
$\overline{\xi}(r, \, t)$, and therefore
we see an enhancement of all the fluctuations.

At small scales the dynamics of the system
shows completely different properties.
For an initial unclustered random distribution
the discrete character of points of the system
dominates the clustering process at small scales.
At the very beginning, i.e. for $t< t_u/4 $,
the fluctuations grow very fast due to
the fall of each particle in the direction of the nearest
particle.  
Actually, Holtsmark distribution 
 shows that the 
 force acting on a particle in an infinite Poisson system
is mainly due to its nearest neighbour.
In this phase the dynamics is therefore dominated by
two-body interactions. 
For times greater than approximately $t_u/4$
and at intermediate scales, the $\Gamma^*(r,\,t)$
 function satisfies the following equation:
\begin{equation}
\Gamma^*( r , \, t + \Delta t ) =
	 \Gamma^*(
		{\rm e}^{-\frac{\Delta t}{\tau_g}} r ,\,
		t
	).
\label{eq:exp_scaling}
\end{equation}
with a measured time-scale $\tau_g = 0.44 \, t_u$, and a similar 
relation also holds for  $\overline{\xi}(r, \, t)$.
This behavior corresponds to a rigid  translation
along the $r$ axis in a logarithmic plot.
This indicates a sort of similarity of structures of different size
at different times. Although it is different from the self-similarity
as usually intended, it is a clear indication of a sort of scale
invariance in the dynamics.
Eq.~(\ref{eq:exp_scaling}) states that the correlations developed
by aggregates of size $r$ at time $t+\Delta t$ are the same as those
developed by aggregates of size $r {\rm e}^{\frac{-\Delta t}{\tau_g}}$ at
time $t$, \emph{and} that the time evolution of such
correlations is the same.

Visual inspection of the evolution of the system (fig.~\ref{fig:projection}) 
suggests that clusters formed at a given time are the new discrete 
elements for the iteration of the clustering process at a larger scale.
We propose a simple  argument which can account for 
the behavior described by Eq.~(\ref{eq:exp_scaling}).
Consider a pair of particles at rest with mass $m_u$ and at distance 
$r_u$, as in the initial condition of our simulations. 
Since the leading contribution of the force acting on a 
particle is mainly due to its nearest neighbor~\cite{hol:17}, 
the time scale $\tau_0$ for their relative motion is:
\begin{equation}
\tau_0 =\sqrt{\frac {r_u^3} {G m_u}}.
\label{eq:timescale_pair}
\end{equation}

After $\Delta t \propto \tau_0$ most of the 
particles are much closer to their nearest neighbor.
We will consider such pairs as "bound systems" approximately 
at rest and distributed randomly with
typical distance $r_1 \approx 2^{1/3} r_u$ and total mass $m_1 = 2 \, m_u$.
We can iterate the same scheme, treating the pairs as the new "particles" 
of the system.
The corresponding time scale $\tau_1$ for the relative motion is:
\begin{equation}
\tau_1 =\sqrt{ {r_1^3}/{G m_1}} = \sqrt{ {2 r_u^3 }/{G 2 m_u }} 
=\tau_0.
\label{eq:timescale_pair2}
\end{equation}
 
Iterating the process we find a sequence of 
``bound systems" with increasing mass $m_n = 2^n \, m_u$ and 
length-scale $r_n = 2^{n/3} r_u$, all with the same
time-scale $\tau_n = \tau_0$.
Therefore, being $n\sim t/\tau_0$, the scale at which this activity 
takes place $r(t)$ 
will obey the following equation:

\begin{equation}
r(t) \propto 2^{\frac{t}{3\tau_0}} r_u
\label{eq:time1}
\end{equation}
which accounts for the exponential dependence of 
Eq.~(\ref{eq:exp_scaling}).

In summary, from the present study we can draw the following conclusions:
(i) we address the issue of the cluster formation  in a gravity driven 
dynamics in a system with random initial positions and
periodic boundary conditions. Strictly speaking, the system has no 
well defined thermodynamic 
limit at asymptotic times. However, it shows a well defined,
thermodynamically stable  clustering in the transient regime;
(ii) we identify the mechanisms of structure formation: initially 
the system tends
to form pairs
These pairs then interact 
with each other and gradually form larger and larger structures, 
{\em shifting the granular structure from small to large scales}.
This process appears to be finally dominating over the fluid like dynamics,
 which tends to amplify the initial large scale fluctuations.
These results  point  to a fundamental importance of the granular
mechanism, which 
appears to lead to a dynamics with elements of self similarity.

To check the validity of our results, we have also 
repeated some of the simulations 
with the GADGET code~\cite{springel:00}.

\acknowledgments
This work was partially supported by the INFM  
under the project {\it Clustering} and by the EC TMR Research Network
under contract ERBFMRXCT960062.

\end{document}